\documentclass[trackchanges,twocolumn]{aastex631}

\usepackage{threeparttable}
\usepackage{longtable}

\newcommand{\Ms}{M_{\odot}}

\newcommand{\iAB}{i_{\mathrm{AB}}}
\newcommand{\MBH}{M_{\mathrm{BH}}}
\newcommand{\Mbulge}{M_{\mathrm{bulge}}}

\newcommand{\Mstar}{M_{\mathrm{star}}}
\newcommand{\Mdyn}{M_{\mathrm{dyn}}}

\newcommand{\Lbol}{L_{\mathrm{bol}}}

\newcounter{num}
\newcommand{\Cfour}{\mathrm{C}\mathrm{\Roman{num}}}
\newcommand{\Mgtwo}{\mathrm{Mg}\mathrm{I}\hspace{-1.2pt}\mathrm{I}}

\newcommand{\Ha}{\mathrm{H}\mathrm{\alpha}}
\newcommand{\Hb}{\mathrm{H}\mathrm{\beta}}

 \newcommand{\mrm}{\ \mathrm}
 \newcommand{\AAA}{\ \mathrm{\mathring{A}}}

\begin{document}

\title{The relationship of SMBHs and host galaxies at $z<4$ in the deep optical variability-selected AGN sample in the COSMOS field}

\author[0009-0003-7134-0539]{Atsushi Hoshi}
    \affiliation{Astronomical Institute, Tohoku University, 6-3 Aramaki, Aoba-ku, Sendai, Miyagi 980-8578, Japan} 
    \affiliation{Institute of Space and Astronautical Science, Japan Aerospace Exploration Agency, 3-1-1, Yoshinodai, Chuou-ku, Sagamihara, Kanagawa, 252-5210, Japan}

\author{Toru Yamada}
\affiliation{Institute of Space and Astronautical Science, Japan Aerospace Exploration Agency, 3-1-1, Yoshinodai, Chuou-ku, Sagamihara, Kanagawa, 252-5210, Japan}
    \affiliation{Astronomical Institute, Tohoku University, 6-3 Aramaki, Aoba-ku, Sendai, Miyagi 980-8578, Japan} 

\author[0000-0001-6402-1415]{Mitsuru Kokubo}
\affiliation{National Astronomical Observatory of Japan, 2-21-1 Osawa, Mitaka, Tokyo 181-0015, Japan} 

\author[0000-0001-5063-0340]{Yoshiki Matsuoka}
\affiliation{Research Center for Space and Cosmic Evolution, Ehime University, 2-5 Bunkyo-cho, Matsuyama, Ehime 790-8577, Japan} 

\author[0000-0002-7402-5441]{Tohru Nagao}
\affiliation{Research Center for Space and Cosmic Evolution, Ehime University, 2-5 Bunkyo-cho, Matsuyama, Ehime 790-8577, Japan} 
\affiliation{Amanogawa Galaxy Astronomy Research Center, Kagoshima University, 1-21-35 Korimoto, Kagoshima 890-0065, Japan}



\begin{abstract}
We present the study on the relationship between SMBHs and their host galaxies using our variability-selected AGN sample ($i_\mathrm{AB} \leq 25.9,\ z \leq 4.5$) constructed from the HSC-SSP Ultra-deep survey in the COSMOS field.
We estimated the BH mass ($M_\mathrm{BH}=10^{5.5-10}\ M_{\odot}$) based on the single-epoch virial method and the total stellar mass ($M_\mathrm{star}=10^{10-12}\ M_{\odot}$) by separating the AGN component with SED fitting.
We found that the redshift evolution of the BH-stellar mass ratio ($M_\mathrm{BH}/M_\mathrm{star}$) depends on the $M_\mathrm{BH}$ which is caused by the no significant correlation between $M_\mathrm{BH}$ and $M_\mathrm{star}$.
Variable AGNs with massive SMBHs ($M_\mathrm{BH}>10^{9}\ M_{\odot}$) at $1.5<z<3$ show considerably higher BH-stellar mass ratios ($>\sim1\%$) than the BH-bulge ratios ($M_\mathrm{BH}/M_\mathrm{bulge}$) observed in the local universe for the same BH range.
This implies that there is a typical growth path of massive SMBHs which is faster than the formation of the bulge component as final products seen in the present day.
For the low-mass SMBHs ($M_\mathrm{BH}<10^{8}\ M_{\odot}$) at $0.5<z<3$, on the other hand, variable AGNs show the similar BH-stellar mass ratios with the local objects ($\sim 0.1\%$) but smaller than those observed at $z > 4$. 
We interpret that host galaxies harboring less massive SMBHs at intermediate redshift have already acquired sufficient stellar mass, although high-z galaxies are still in the early stage of galaxy formation relative to those at the intermediate/local universe.

\end{abstract}

\section{Introduction}


It is widely accepted that Supermassive Black Holes (SMBHs) are located at the center of their host galaxies.
The mass of SMBHs ($\MBH$) and the bulge mass of their hosts ($\Mbulge$) show a well-established tight correlation in the local universe \citep[e.g.,][]{magorrian1998demography,haring_black_2004,kormendy2013coevolution,mcconnell2013revisiting}. 
This correlation indicates that the growth of SMBHs and the formation of galaxies are tightly related to each other.
There is also a scaling relation between $\MBH$ and the total stellar mass ($\Mstar$), which is more scattered compared to the $\MBH-\Mbulge$ relation \citep[e.g.,][]{beifiori2012correlations,kormendy2013coevolution,reines2015relations,davis2018blackspiral}.
The scatter of the scaling relation depends on the morphology of the host galaxies \citep[e.g.,][]{davis2018blackspiral,sahu2019blacketgs}.
Since $\Mstar$ includes not only the bulge but the disk component of the host galaxy, the scaling relation is known to shift toward the direction of larger galaxy mass compared to the $\MBH-\Mbulge$ relation.
For instance, spiral galaxies generally possess larger stellar masses than elliptical or S0 galaxies when compared at the same BH mass range. 

To understand the origin of the local $\MBH-\Mbulge$ relation, it's essential to explore the relationship between SMBHs and galaxies not just in the nearby universe but also at the intermediate (e.g., $0<z<4$) and high redshift (e.g., $4<z$).
Since it is challenging to estimate $M_{\mathrm{bulge}}$ accurately beyond the local universe, the stellar mass $M_\mathrm{star}$ or the dynamical mass $M_\mathrm{dyn}$ has been used instead of the bulge mass $M_{\mathrm{bulge}}$.
At intermediate redshift, \citet{decarli10} shows that the BH-stellar mass ratio ($\MBH/\Mstar$) increases by a factor of $\sim 7$ from $z=0$ to $z=3$, \citep[see also][]{merloni2009cosmic}, while \citet{suh2020no} shows that there is no significant redshift evolution of the BH-stellar mass ratio up to $z=2.5$.
At high redshift, some studies \citep[e.g.][]{targett2012host,ding2020mass,pensabene2020alma,harikane2023jwst} show that the high BH-stellar mass ratios (or the high BH-dynamical mass ratios) tend to be observed. It is interpreted that SMBHs grow faster than their host galaxies, although the selection bias for luminous objects should be carefully considered.
Indeed, the Subaru high-z exploration of low-luminosity quasars (SHELLQs) revealed that relatively low-luminosity quasars  ($M_{1450} > -25$ mag) at $z>6$ seem to be consistent with the local relation, particularly at the high dynamical mass range of $\Mdyn>4\times10^{10}\ \Ms$ \citep{izumi2019subaru,izumi2021subaru}.
To avoid such possible bias, it is necessary to construct a sample of AGN covering the objects with sufficiently low luminosity. 
However, constructing a less biased AGN sample is difficult due to the limitations of each selection method.
Optical and IR color selection methods may be affected by the host galaxy contamination for low-luminosity AGN \citep[e.g.,][]{boutsia2009spectroscopic}.
X-ray searches for the faint or distant objects are limited by the sensitivity of the current facilities.
Emission line diagnostics \citep[e.g.,][]{bptdiagram} are also affected by contamination from the host galaxies. 


In this paper, we aim to explore the relationship between SMBHs and galaxies utilizing a variability-selected AGNs sample ($i_\mathrm{AB} \leq 25.9,\ z \leq 4.5$) constructed by \citet{kimura_properties_2020}.
The optical variability-based AGN selection method has advantages in exploring the low-mass SMBH beyond the local universe.
This method allows only Type1 AGNs to be selected independent of the broad line width in principle, as the central variable source inside BLR is not obscured by dust torus. 
Consequently, there's no need to set a threshold for the minimum velocity width (e.g., $\Delta V=2000\ \mathrm{km\ s^{-1}}$, \citep{merloni2009cosmic,suh2020no}) to distinguish between Type 1 and Type 2 AGNs.
Moreover, since there is a strong anti-correlation between variability amplitude and AGN luminosity \citep[e.g.,][]{berk2004ensemble}, the deep optical variability-based AGN selection can construct the low luminosity Type1 AGN sample \citep[e.g.,][]{kimura_properties_2020,guo2020dark}.
The structure of the paper is as follows. In section 2, we explain the sample of variability-selected AGNs and the photometry and the spectroscopic data to estimate the BH mass and the stellar mass. In section 3, we describe the single epoch virial mass method to estimate the BH mass and the spectral energy distribution (SED) fitting model to measure the stellar mass. In sections 4 and 5, we present the results and discussion on the relationship of SMBHs and the galaxies using the variability-selected AGN sample constructed from the HSC-SSP survey.   
Throughout this paper, we assume a $\mathrm{\Lambda CDM}$ cosmological parameters of $\Omega_m=0.3$ and $\Omega_\Lambda =0.7$ and Hubble constant $H_0=70\mrm{km\ s^{-1} Mpc^{-1}}$. We use the AB magnitude system.
\section{Dataset}
\subsection{Variability-based Sample}
The Hyper Suprime-Cam Subaru Strategic Program (HSC-SSP) survey consists of three layers (Wide: 1400 deg, Deep: 27 deg, Ultradeep: 3.5 deg) with multi-band ($g,\ r,\ i,\ z,\ y$-band and 4 narrow band) filters, achieving a deep multi-band optical imaging observation with the limiting magnitude $\iAB \sim 27.4$ in the Ultradeep layer \citep{aihara2018first,aihara2018hyper}.
Using the data set of HSC-SSP survey in the Ultradeep layer in the COSMOS field, \citet{kimura_properties_2020} conducted a variability search to detect the 491 robust AGNs.
This parent sample covers the redshift up to $z=4.5$ and the $i$-band AB magnitude down to $i_\mathrm{AB}=25.9$ including 441 X-ray detected AGNs and 50 X-ray undetected AGNs.
In this paper, 'X-det varAGNs' ('X-undet varAGNs') refer to objects that are detected as X-ray sources by the Chandra COSMOS Legacy Survey \citep{laigle2016cosmos2015, marchesi2016chandra} with a signal-to-noise ratio (S/N) greater (less) than 3.

\subsection{Photometry and Spectroscopy}
We used the latest UV-IR photometry dataset of COSMOS2020 \citep{weaver2022cosmos2020} and the deep X-ray catalog of Chandra COSMOS Legacy Survey \citep{laigle2016cosmos2015,marchesi2016chandra} to obtain the SED of varAGNs
The X-ray flux limits for $20\%$ completeness correspond to $1.3\times10^{-15}\ \mathrm{erg\ s^{-1}\ cm^{-2}}$ for the full band (0.5-10 keV), $3.2\times10^{-16}\ \mathrm{erg\ s^{-1}\ cm^{-2}}$ for the soft band (0.5-2 keV), and $2.1\times10^{-15}\ \mathrm{erg\ s^{-1}\ cm^{-2}}$ for the hard band (2-10 keV), respectively.
To estimate the BH mass, we compiled the following archival spectroscopic dataset.
From SDSS DR16, we use the data with a spectral resolution of $R\sim2000$ covering the wavelength range $3600\AAA<\lambda<10400\AAA$ \citep{ahumada202016th}. 
We also use the data from Keck DEIMOS with $R\sim6000$ covering $4100\AAA<\lambda<11000\AAA$ \citep{hasinger2018deimos}.
zCOSMOS provides the data with $R\sim500$ within the range $5500\AAA<\lambda<9500\AAA$ \citep{lilly2007zcosmos}.
FMOS-COSMOS version 2.0 provides near-infrared spectroscopic data at resolutions of $R\sim600\ \mathrm{or}\ 3000$ covering $11100\AAA<\lambda<18000\AAA$ \citep{silverman2015fmos,kashino2019fmos}. Finally, we also use the data obtained with the Dark Energy Spectroscopic Instrument (DESI) at resolutions of $2000< R< 5500$ covering $3500\AAA<\lambda<10000\AAA$  \citep{desi2022overview,desi2023bedr}.
In total, we estimated $\MBH$ of the 208 objects using the available spectroscopic data, however, we were unable to determine the broad line width for 283 objects, especially for the faint objects.
The redshift distribution of AGNs for which we measured the broad line width is similar to that of the parent sample.
Note that the requirement for sufficient S/N in spectroscopic data for broad line width measurements could bias the sample towards more luminous and massive objects.
We anticipate uncovering further low-mass SMBHs through the deeper spectroscopic observations in the future. 
\section{Method}
\subsection{SMBH mass estimate}
\begin{figure*}[htbp]
    
    \begin{tabular}{cc}
      \begin{minipage}[t]{0.42\hsize}
        \centering
        \includegraphics[keepaspectratio, scale=0.4]{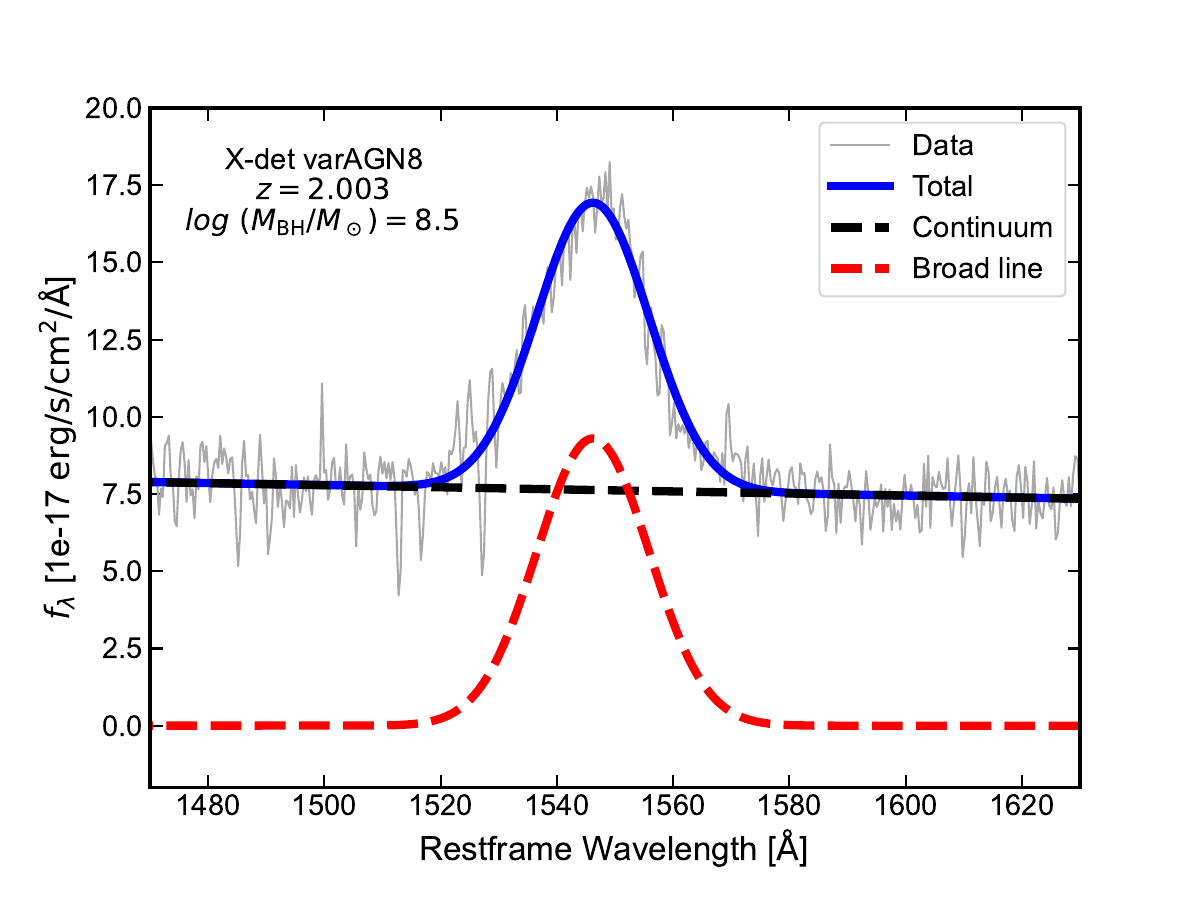}
      \end{minipage} &
      \begin{minipage}[t]{0.42\hsize}
        \centering
        \includegraphics[keepaspectratio, scale=0.4]{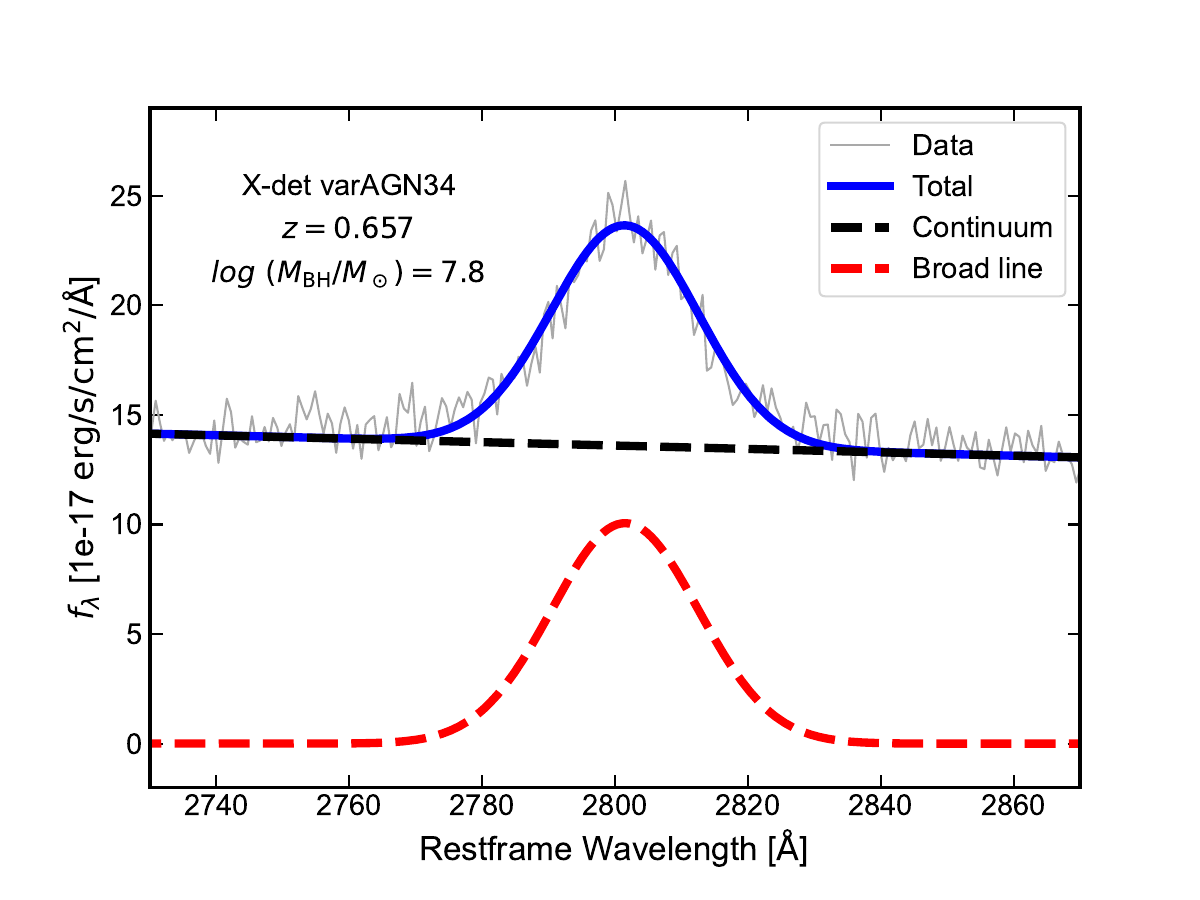}
      \end{minipage} \\
      \begin{minipage}[t]{0.42\hsize}
        \centering
        \includegraphics[keepaspectratio, scale=0.4]{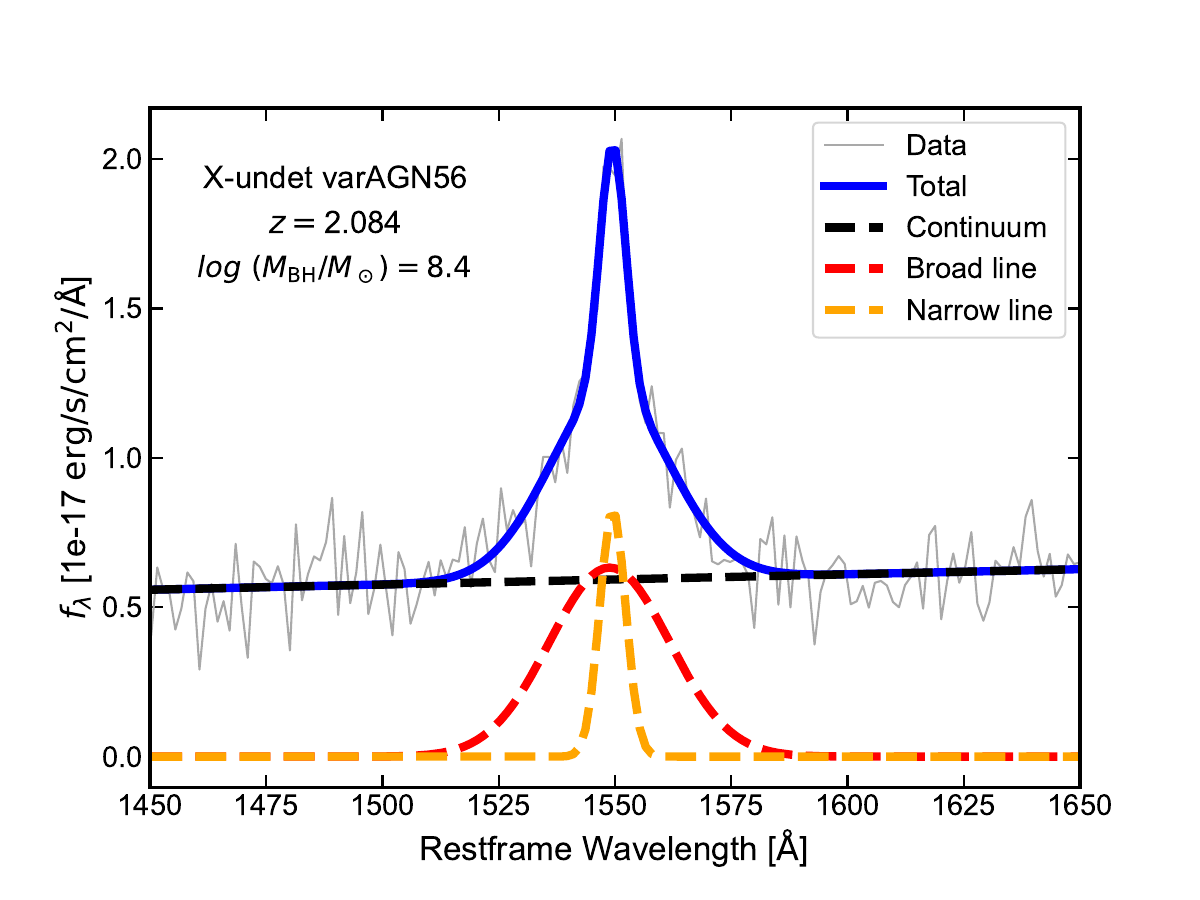}
      \end{minipage} &
      \begin{minipage}[t]{0.42\hsize}
        \centering
        \includegraphics[keepaspectratio, scale=0.4]{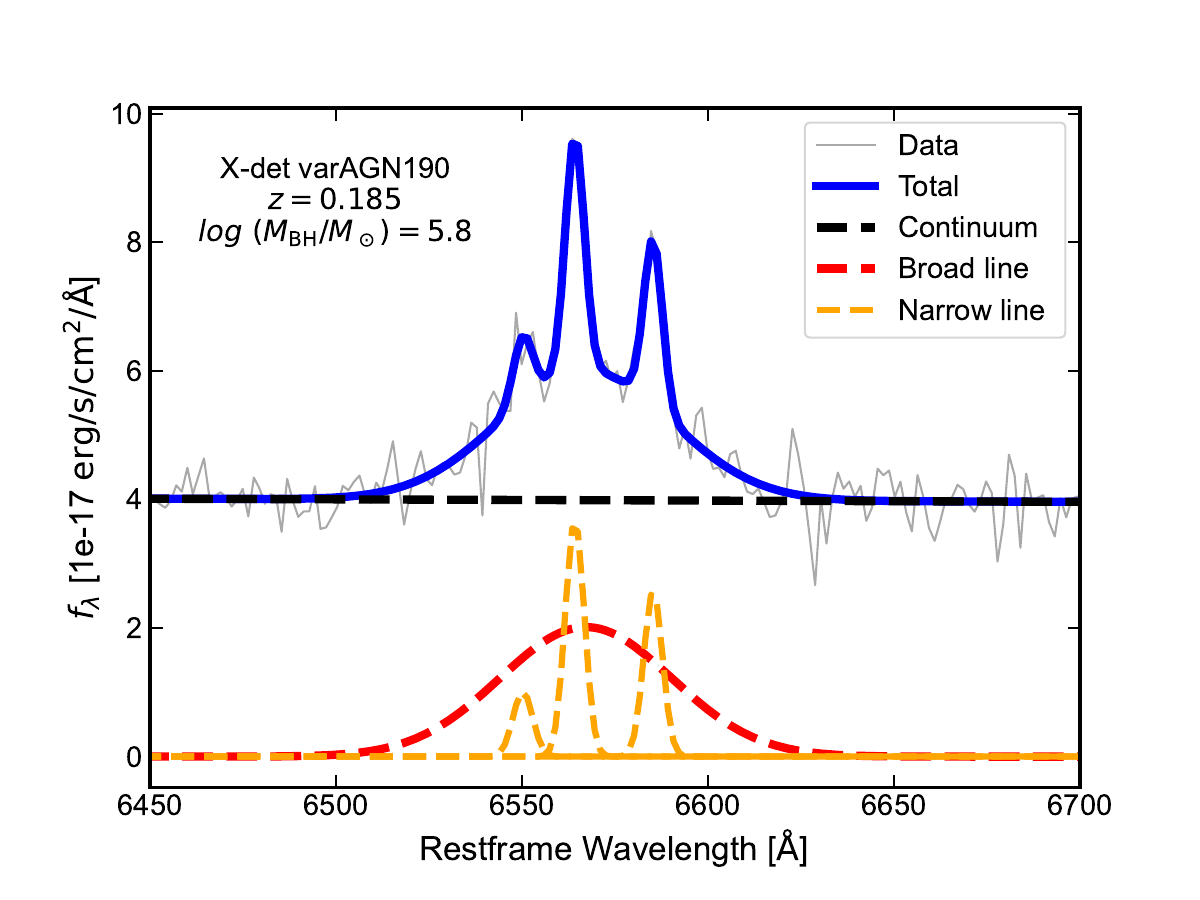}
      \end{minipage} \\
      \begin{minipage}[t]{0.42\hsize}
        \centering
        \includegraphics[keepaspectratio, scale=0.4]{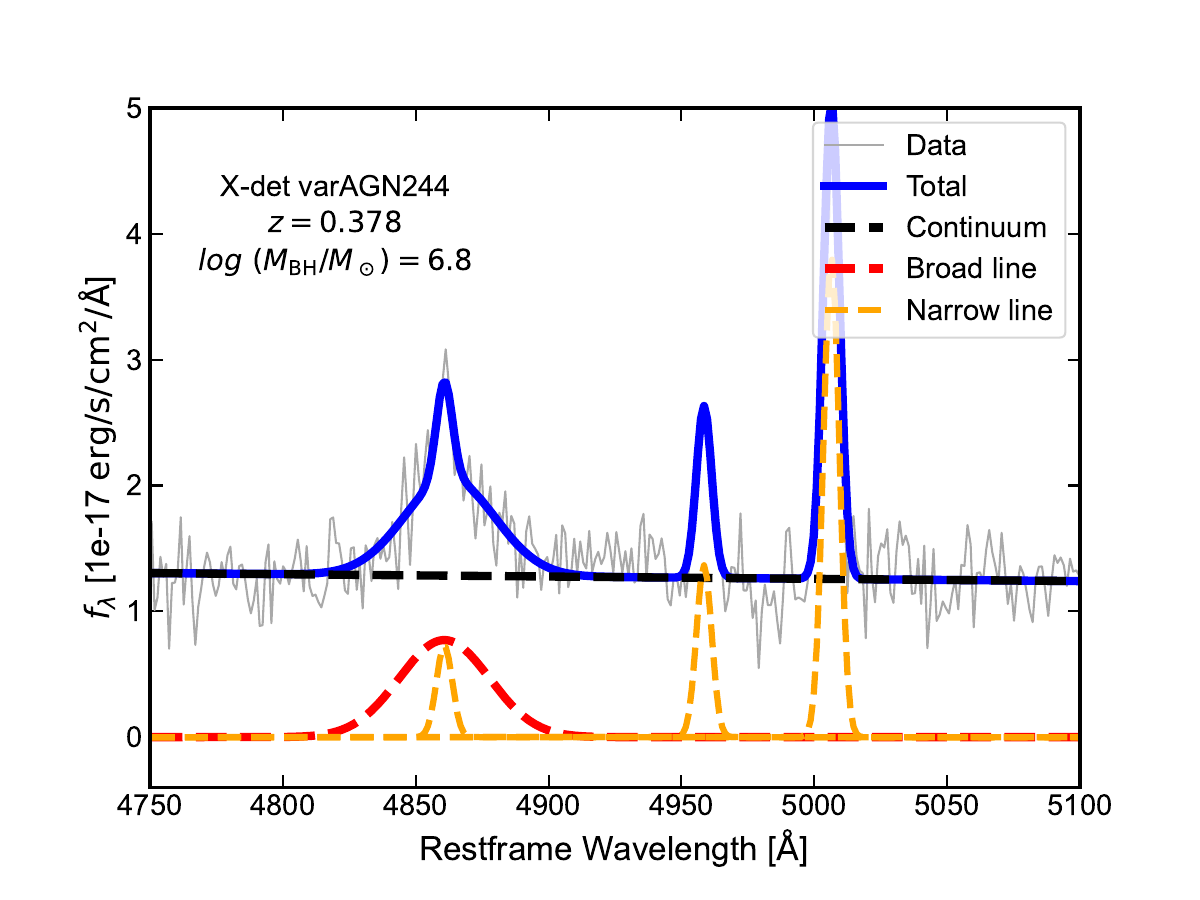}
      \end{minipage} &
       \begin{minipage}[t]{0.42\hsize}
        \centering
        \includegraphics[keepaspectratio, scale=0.4]{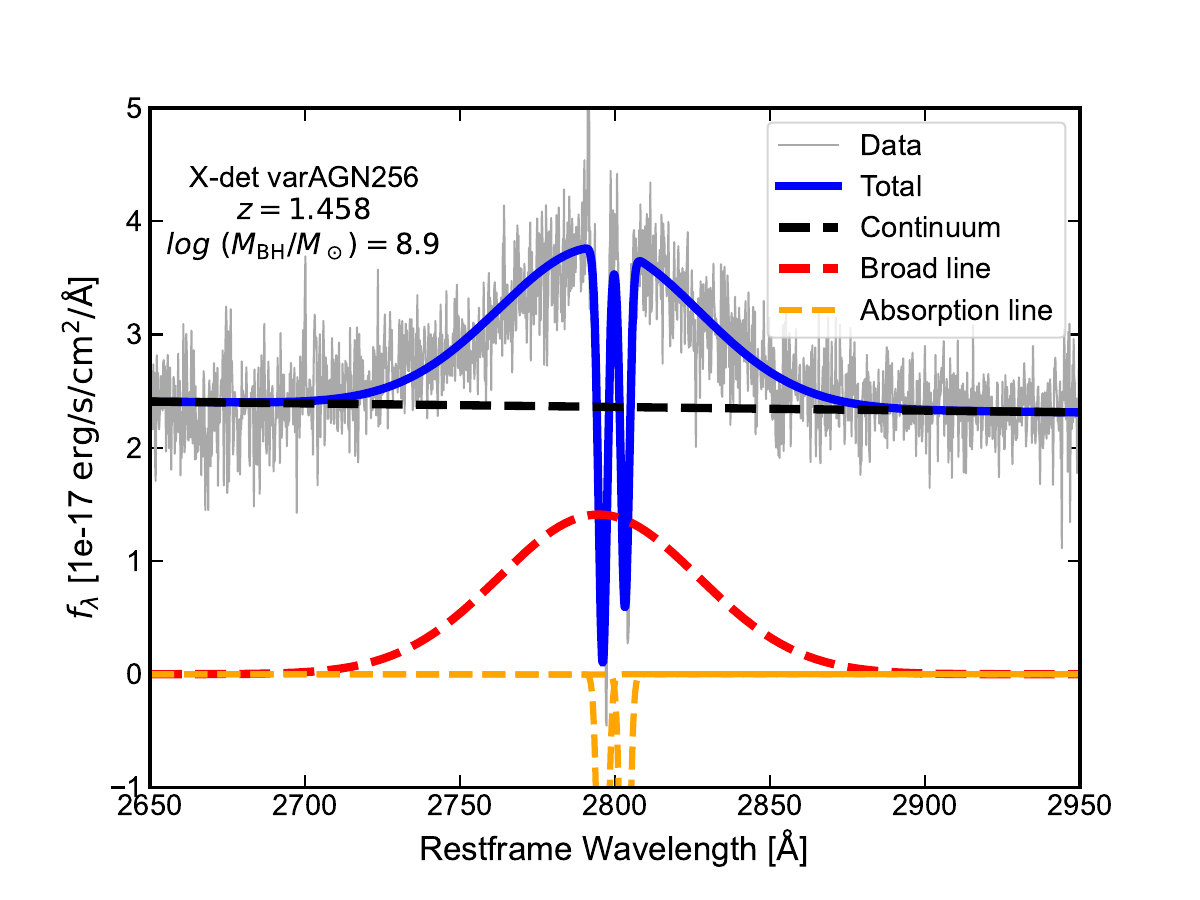}
      \end{minipage} \\
    \end{tabular}
    \caption{Six typical examples of spectral fitting around the broad line for varAGNs. Gray and blue lines show the restframe observed spectra and the total fitting results respectively. Red, orange and black dashed lines represent the fitting component of the broad line, the narrow (or the absorption) lines and the continuum respectively. Each varAGN ID taken from \citet{kimura_properties_2020}, the spectroscopic redshift, and the BH mass are displayed in upper left area of each panel.}\label{six-example}
  \end{figure*}

Since the optical variability-selected AGNs can be generally classified as Type1 AGN, independent of their broad line velocity width, we utilize the single epoch virial method to estimate their black hole mass.
Assuming the material in BLR is under the virial equilibrium, we obtained the black hole mass following the equation,

$\log(\MBH/\Ms)= \\6.66+ 0.53\log \left(\frac{\lambda L_{\lambda(1350)}}{10^{44}\ \mathrm{erg\ s^{-1}}}\right)+2\log \left(\frac{FWHM_{\Cfour}}{10^3\ \mathrm{km\ s^{-1}}}\right)\\
6.79+ 0.5\log \left(\frac{\lambda L_{\lambda(3000)}}{10^{44}\ \mathrm{erg\ s^{-1}}}\right)+2\log \left(\frac{FWHM_{\Mgtwo}}{10^3\ \mathrm{km\ s^{-1}}}\right),\\
6.64+ 0.64\log \left(\frac{\lambda L_{\lambda(5100)}}{10^{44}\ \mathrm{erg\ s^{-1}}}\right)+2\log \left(\frac{FWHM_{\Hb}}{10^3\ \mathrm{km\ s^{-1}}}\right),\\
6.70+ 0.64\log \left(\frac{\lambda L_{\lambda(5100)}}{10^{44}\ \mathrm{erg\ s^{-1}}}\right)+2.06\log \left(\frac{FWHM_{\Ha}}{10^3\ \mathrm{km\ s^{-1}}}\right)$\\
where $\lambda L_\lambda$ is the AGN luminosity $\mathrm{[erg\ s^{-1}]}$ at the wavelength $\lambda$, $FWHM_\mathrm{line}$ is the full width at half maximum for $\Cfour,\ \Mgtwo,\ \Hb,\ \mathrm{and}\ \Ha$ broad emission lines \citep{vestergaard2006c4,bahk2019calibrating,greene2005estimating}.
For the objects with multiple broad emission lines detected, we selected lines from spectroscopic data with higher wavelength resolution or higher S/N ratios.
The reliability of mass measurements obtained with $\Cfour$ line has been studied \citep[e.g.,][]{coatman2016c,marziani2019black}. While luminous AGNs with the higher bolometric luminosity ($\Lbol > 10^{46.5}$) tends to have larger differences in BH mass measured by $\Cfour$ and $\Hb$, most of our sample has a moderate bolometric luminosity ($\Lbol= 10^{44-46}$).
To obtain $\lambda L_{\lambda (5100)}$ for X-det varAGNs, we applied the bolometric correction from the hard X-ray luminosity ($BC_\mathrm{X}=\Lbol / \lambda L_{\lambda(\mathrm{2-10\ keV})}$) represented by the polynomial equation as a function of bolometric luminosity \citep{lusso2012bolometric} and the optical conversion factor ($BC_\mathrm{O}=\Lbol / \lambda L_{\lambda (5100)}$) of 10.33 \citep{richards2006spectral}.
The optical slope $\alpha _{\mathrm{o}}=-0.5$ is used to convert from the $\lambda L_{\lambda (5100)}$ to $\lambda L_{\lambda (3000)}$ or $\lambda L_{\lambda (1350)}$ \citep{richards2006spectral}.
For the X-undet varAGNs with $S/N<3$ or no detection in hard X-ray, the AGN bolometric luminosity is calculated from the parameter {\tt\string accretion luminosity} in the SED fitting results.
We will explain the SED fitting method in $\S$ \ref{host}.
For the measurement of $FWHM_\mathrm{line}$, we perform spectral fitting utilizing the {\tt\string curve\_fit} tool from Python, a function from the {\tt\string scipy} module \citep{virtanen2020scipy}.
We fit the continuum with a linear component and the broad line with a single Gaussian component.
If the spectrum around the broad line shows narrow lines or absorption features, additional Gaussian components are incorporated into the fitting model. 
We visually inspected carefully all the spectra to remove objects with poor fitting accuracy even if the broad lines are detected.
Since optical variability-selected AGNs are essentially classified as Type 1 AGNs, the minimum threshold for $FWHM_\mathrm{line}$ is set as small as $500\ \mathrm{km\ s^{-1}}$.
We also excluded objects where the line width of $\Cfour,\ \Mgtwo,\ \Hb,\ \mathrm{and}\ \Ha $ is similar with those of the forbidden lines in the spectra and difficult to identify the broad line components.
Very low-mass SMBHs with the velocity width less than $500\ \mathrm{km\ s^{-1}}$ are thus not included in our sample. 
We show the six typical examples of the spectra to obtain SMBH mass by using the single epoch virial method in Figure \ref{six-example}.

\subsection{Host galaxies}\label{host}

To estimate the AGN luminosity and the stellar mass, we utilize the SED fitting code CIGALE-v2022 \citep{boquien2019cigale}.
We use the multi-wavelength photometry catalog of COSMOS2020 \citep{weaver2022cosmos2020} and the X-ray photometry catalog of Chandra COSMOS Legacy survey \citep{marchesi2016chandra} and the filter band used in the fitting is summarized in Table \ref{sed-filter}.
We adopt the delayed star formation history \citep{ciesla2017sfr} and the stellar synthesis population \citep{bruzual2003stellar} with a Salpeter initial mass function \citep{salpeter1955luminosity} for stellar SED models. 
The nebular emission from the photoionization models \citep{inoue2011rest} is included. 
We use the dust attenuation in the interstellar medium \citep{charlot2000simple} and the heated dust emission \citep{dale2014two}.
The  SKIRTOR AGN model \citep{stalevski20123d,stalevski2016dust} is used for the AGN SED model.
The photon index of the X-ray \citep{stalevski2016dust} takes into account not only Type 1 AGN but also the hard spectrum seen in Type 2 AGN since there are some varAGNs that are strongly absorbed by (soft) X-ray \citep{kimura_properties_2020}.
We performed SED fitting without using soft-X-ray data for the varAGN which shows strong X-ray absorption with X-ray hardness ratio $HR>0$.
All the SED fitting parameters are listed in Table \ref{CIGALE model}.
We set the criteria of the reduced chi-square $\chi^2_\mathrm{red}<5$ for varAGNs.
\section{Results}
\subsection{Redshift distribution of the BH mass}
\begin{figure}[htbp]
\begin{center}
\includegraphics[width=80mm]{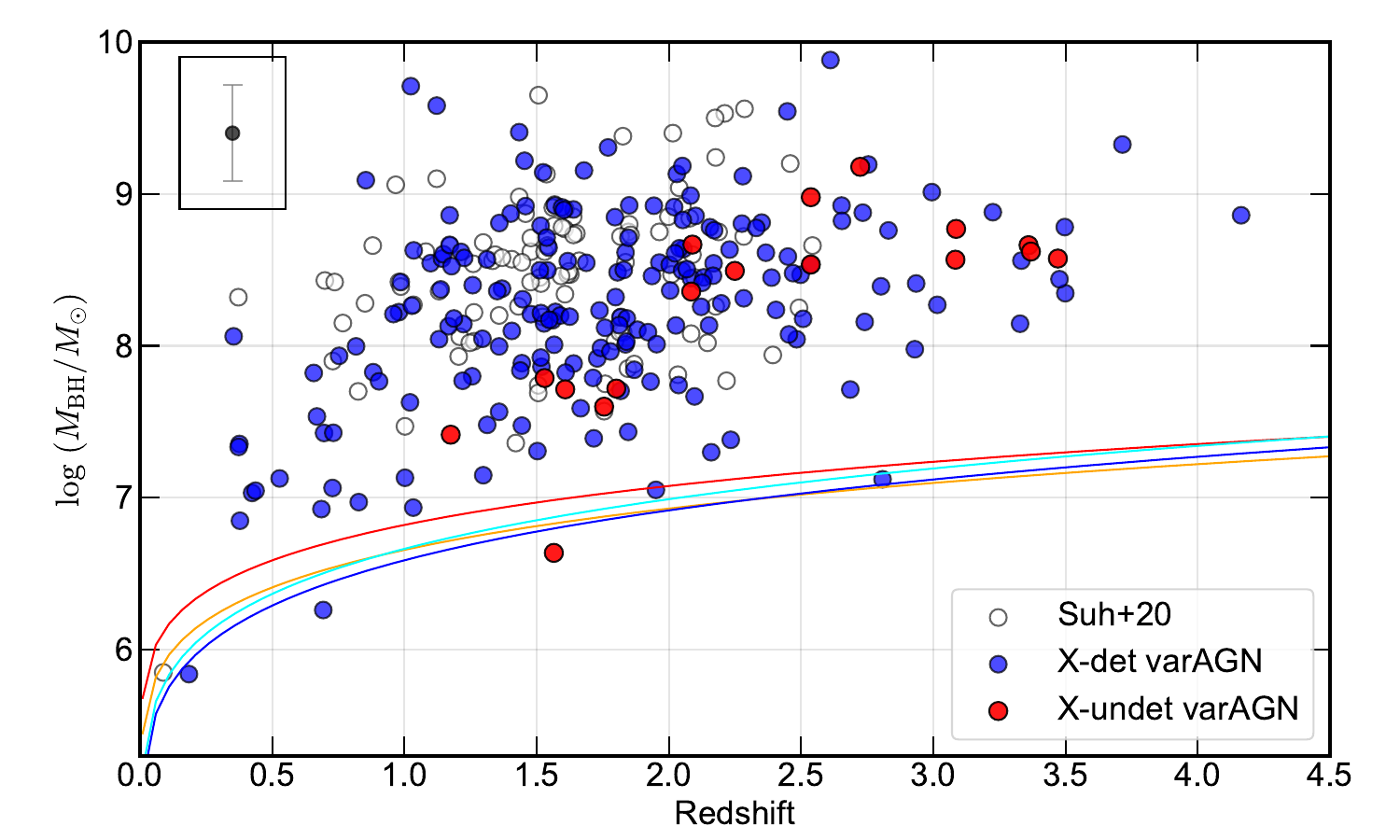}
\caption{The distribution of $\MBH$ as a function of redshift. Blue and red circles represent X-det and X-undet varAGN samples. We plot X-ray detected AGNs which have the broad line velocity width $\Delta V >2000\ \mathrm{km\ s^{-1}}$ estimated by \citet{suh2020no} shown by black open circles.  The orange, red, blue and cyan solid line show the detection limits of black hole mass as a function of redshift derived from the $20\%$ completeness of hard X-ray and the velocity width $\Delta V =2000\ \mathrm{km\ s^{-1}}$ in $\Cfour,\ \Mgtwo,\ \Hb\ \mathrm{and}\ \Ha$ respectively. The black filled circle and error bar at the upper left represent the mean error for the X-det varAGN sample.}
\label{z-mass}
\end{center}
\end{figure}

We showed the obtained SMBH mass for the 191 X-det varAGNs and the 17 X-undet varAGNs in Figure \ref{z-mass} and the results are also summarized in Table \ref{bhmass-table1}.
We identified 6(7) varAGNs with $10^{6(5)}\leq\MBH/\Ms<10^7 $ up to $z\sim1.6$,
53 varAGNs with $10^7\leq\MBH/\Ms<10^8$ up to $z\sim2.9$,
131 varAGNs with $10^8\leq\MBH/\Ms<10^9$ up to $z\sim4.2$, and
17 varAGNs with $10^9\leq\MBH/\Ms<10^{10}$ up to $z\sim3.7$.
We measured BH mass by $\Cfour$ line for 63 objects, by $\Mgtwo$ for 127 objects, by $\Hb$ for 9 objects, and by $\Ha$ for 9 objects. 
The colored solid lines show the detection limit of the low mass $\MBH$ as a function of redshift, which refers the 20$\%$ completeness limit which corresponds to the hard band X-ray of $f_\mathrm{X} =2.5\times 10^{-15}\ \mathrm{erg\ s^{-1}\ cm^{-2}}$ (2-7 keV, $\alpha_\mathrm{x}=-1.5$, \citet{marchesi2016chandra}) and the broad line velocity width of $\Delta V=2000\ \mathrm{km\ s^{-1}}$ \citep[e.g.][]{merloni2009cosmic,suh2020no}).
Notably, 5 varAGNs are plotted below this relatively conservative X-ray detection limit, which shows that the optical variability-selected method is an effective tool for detecting low mass SMBHs beyond the local universe.
While the majority of AGNs at $z>1$ in literature are limited to SMBHs of $\MBH>10^8\ \Ms$ \citep{shen2011catalog}, our sample is unique in encompassing low mass SMBHs of $\MBH<10^8\ \Ms$ extending to $z=3$.
Furthermore, massive SMBHs with $\MBH>10^{9}\ \Ms$ are also included in our sample.
Regarding X-undet varAGNs, the SMBH masses are lower than those of the X-det varAGN at $z\sim2$.
Our sample covers a wide range of redshifts ($0<z<4.2$) and SMBH mass ($5.5<\mathrm{log}\ (\MBH/\Ms)<10$).


\subsection{Eddington Ratio and the BH mass}
\begin{figure}[htbp]
\begin{center}
\includegraphics[width=80mm]{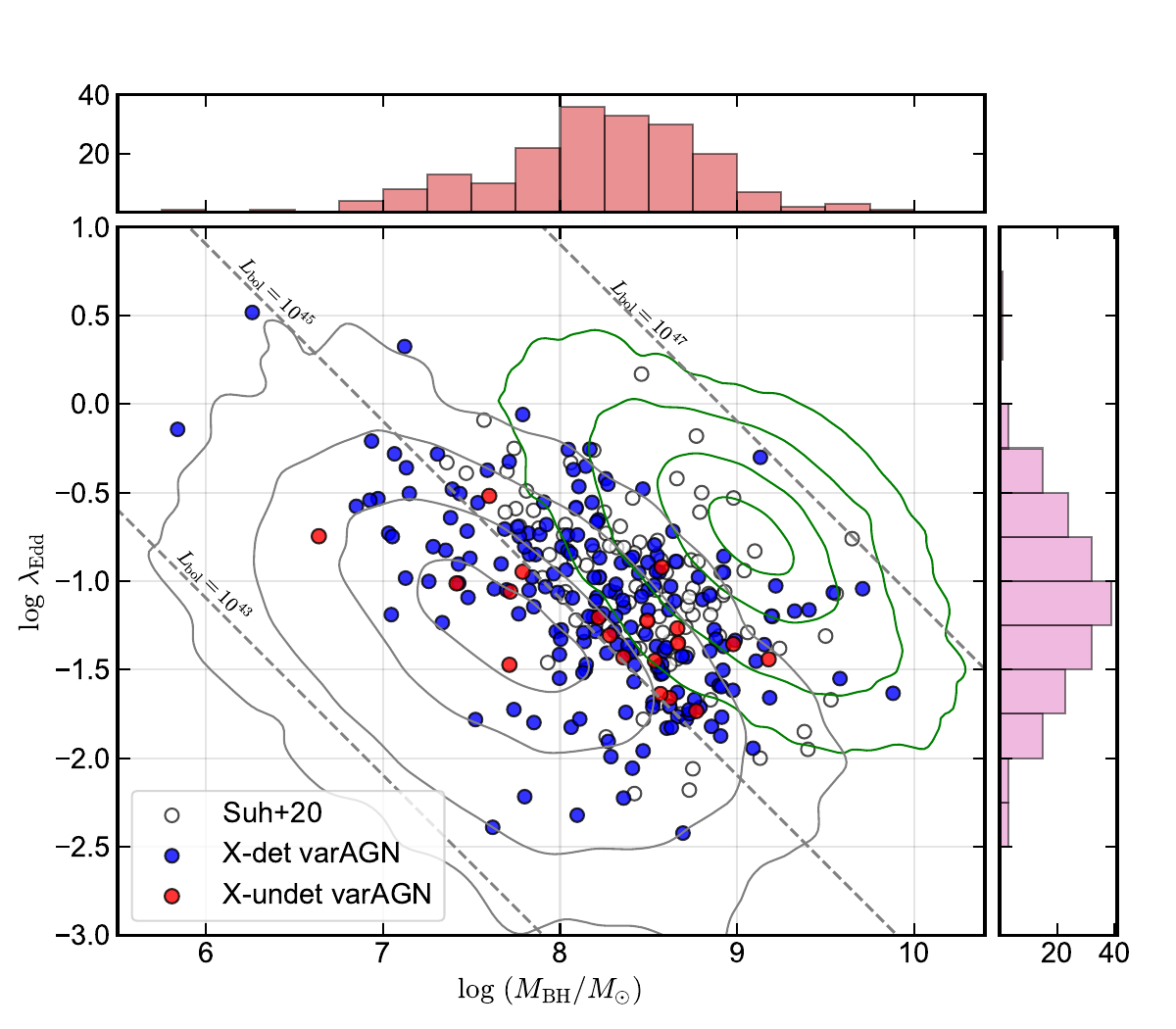}
\caption{The relationship between BH mass and Eddington ratio. Blue and red circles show X-det varAGNs and X-undet varAGNs respectively. Black open circles show X-detected broad line AGNs from \citet{suh2020no}. Gray and green contours represent the relation at $z<0.35$ \citep{liu2019comprehensive} and $1\leq z<2$ \citep{shen2011catalog} respectively. The dashed lines represent the bolometric luminosity of $\Lbol =10^{43},\ 10^{45},\ 10^{47}\ \mathrm{erg\ s^{-1}}$. The top and right histograms represent the distribution of BH mass and Eddington ratio in our X-det varAGN sample, respectively.}
\label{edd-mass}
\end{center}
\end{figure}

We show the relationship between BH mass and Eddington ratio ($\lambda_\mathrm{Edd}$) in varAGN sample in Figure \ref{edd-mass}. 
Many of our varAGN sample shows the Eddington ratio around $\lambda_\mathrm{Edd}=0.1$ which is consistent with the properties of the typical Type1 AGN \citep[e.g.][]{caccianiga2013black}.
The mean Eddington ratio in X-det and X-undet varAGN sample are $0.10,\ \mathrm{and}\ 0.06$ respectively.
X-undet varAGNs show slightly lower accretion rates than X-det varAGNs.
The gray and green contours represent the distribution of the objects in literature at $z<0.35$ \citep{liu2019comprehensive} and $1\leq z<2$ \citep{shen2011catalog}.
Our variability sample ($0<z<4.2$) is located between these samples at $z<0.35$ and $1\leq z<2$ due to the difference in the observed volume and redshift.

\subsection{The relation of SMBHs and host galaxies}
\begin{figure}[htbp]
\begin{center}
\includegraphics[width=80mm]{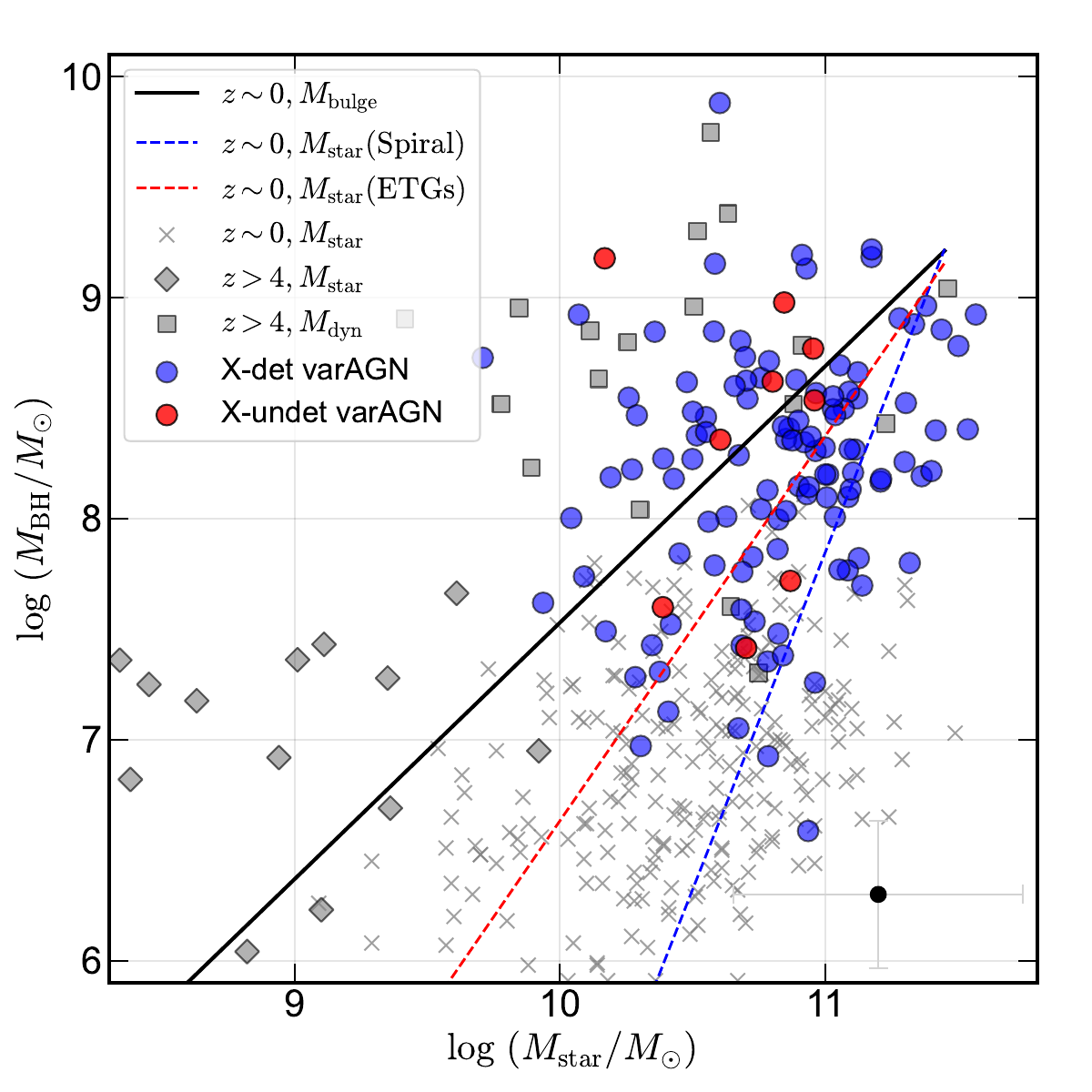}
\caption{The relationship of $\MBH$ and $\Mstar$ in X-det varAGN and X-undet varAGN shown by blue and red circles respectively. The black circle and error bar at the lower right represent the mean error of the X-det varAGN sample. The black solid line shows $\MBH-\Mbulge$ relation for elliptical galaxies and bulge dominant galaxies in the local universe \citep{kormendy2013coevolution}. The red and blue dashed lines represent $\MBH-\Mstar$ relation in the local universe for early type galaxies (ETGs, \citep{sahu2019blacketgs}) and for spiral galaxies (Spiral, \citep{davis2018blackspiral}) respectively. The gray cross markers show the $\MBH-\Mstar$ relation in AGN sample at the nearby universe \citep{reines2015relations}. The gray diamonds show $\MBH-\Mstar$ in AGN sample at $z\sim4$ \citep{harikane2023jwst,maiolino2023jades} from JWST. The gray squares show $\MBH-\Mdyn$ in AGN sample at $z=4\sim7$ from ALMA \citep{izumi2019subaru,pensabene2020alma,izumi2021subaru}.}
\label{bh-st}
\end{center}
\end{figure}

The $\MBH-\Mstar$ relation in our varAGN sample is shown in Figure \ref{bh-st} and the results are listed in Table \ref{bh-ste-table}.
As we plotted the objects whose stellar mass can be obtained by the SED model fitting (Section 2), the number of objects displayed in Figure \ref{bh-st} is limited to 111 X-det varAGNs and 9 X-undet varAGNs.
As our spectroscopic sample is limited by the available archived data, we infer that the completeness is low at the low-mass region in both $\MBH <10^{7}\ \Ms$ and $\Mstar <10^{10}\ \Ms$.
Since there seems no significant difference in the distribution X-det and X-undet varAGNs, we do not treat them separately in the following discussion and hereafter simply refer them as varAGN.
For the reference, $\MBH$ and $\Mstar$ for AGNs in the local universe ($z<0.055$) are shown by gray crosses \citep{reines2015relations}.
The black solid line represents $\MBH-\Mbulge$ relation for the elliptical galaxies and bulge dominant galaxies in the local universe \citep{kormendy2013coevolution}.
The red and blue dashed lines represent $\MBH-\Mstar$ relation in the local universe for early type galaxies (ETGs, \citep{sahu2019blacketgs}) and for spiral galaxies (Spiral, \citep{davis2018blackspiral}) respectively.
We also plot the high-z AGNs detected by JWST \citep[e.g.,][]{harikane2023jwst,maiolino2023jades} and by ALMA \citep{izumi2019subaru,pensabene2020alma,izumi2021subaru} by gray diamonds and gray squares, respectively.

Our sample is widely scattered and appears to be no strong correlation between $\MBH$ and $\Mstar$.
The correlation coefficient is $0.196$.
For varAGNs with smaller $\MBH$ relative to the $\MBH-\Mbulge$ relation in the local universe, these BH-stellar mass ratios seem to be consistent with the local ETGs, and spiral galaxies.
Furthermore, when comparing AGNs at $z<0.055$ (gray crosses) with varAGNs in the low-mass SMBH range ($\MBH<10^8\ \Ms$), there appears no significant difference.
The stellar mass of these objects at intermediate redshift is also likely to be affected by the disk components although their true morphology is yet to be investigated.
This will be discussed in Section 5.

On the other hand, we found that varAGNs with significantly massive SMBH compared to $\MBH-\Mbulge$ relation in the local universe \textbf{are} hard to explain by the disk component.
Overmassive AGNs are frequently observed at high redshift (gray diamonds and gray squares), and their presence is seen among varAGNs even at intermediate redshift.
This implies that there is a typical growth path of massive SMBHs which is faster than the formation of their host galaxies.

\begin{figure*}[htbp]
\begin{center}
\includegraphics[width=180mm]{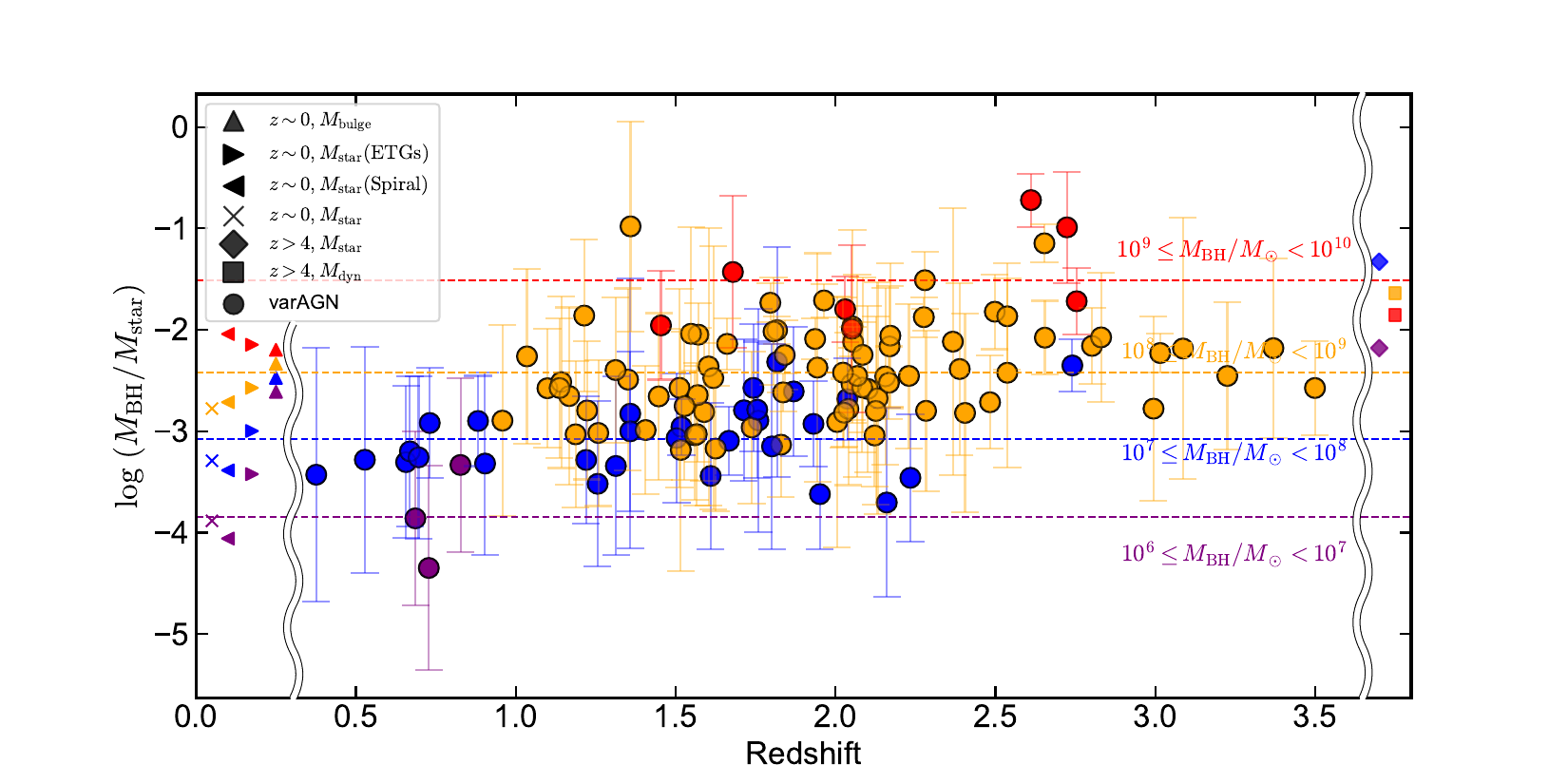}
\caption{The redshift evolution of the BH-stellar mass ratio : X-det and X-undet varAGNs are shown by the filled color circles and the colors represent different BH mass ranges. The dashed colored lines represent the average BH-stellar mass ratio within each BH mass range of varAGN. 
Crosses, left triangles and right triangles at the left side of the panel show the BH-stellar mass ratios with $\mathrm{log}\ (\MBH/\Ms)={6.5, 7.5, 8.5, 9.5}$ of AGNs \citep{reines2015relations}, spiral galaxies \citep{davis2018blackspiral} and ETGs \citep{sahu2019blacketgs} in the local universe, respectively. 
Upward triangles show the BH-bulge mass ratios with $\mathrm{log}\ (\MBH/\Ms)={6.5, 7.5, 8.5, 9.5}$ of elliptical and S0 galaxies \citep{kormendy2013coevolution}.
Squares and the diamonds show the average BH-dynamical mass \citep{izumi2019subaru,pensabene2020alma,izumi2021subaru} with the BH mass range of $\mathrm{log}\ (\MBH/\Ms)>{8}$ and the average BH-stellar mass ratio \citep{harikane2023jwst,maiolino2023jades} with the BH mass range of $\mathrm{log}\ (\MBH/\Ms)<{8}$ at high-z.
}
\label{s-ratio-bin}
\end{center}
\end{figure*}
\subsection{The BH-stellar mass ratio as a function of the redshift}
Figure \ref{s-ratio-bin} shows the BH-stellar mass ratio ($\MBH/\Mstar$) in our sample as a function of redshift and the comparison with the different redshift samples.
Colors represent different BH mass ranges, namely, purple, blue, orange, and red correspond to BH mass ranges of $10^6\leq\MBH/\Ms<10^7$, $10^7\leq\MBH/\Ms<10^8$, $10^8\leq\MBH/\Ms<10^9$, and $10^9\leq\MBH/\Ms<10^{10}$ respectively.
The dashed colored lines represent the average BH-stellar mass ratio $-3.9\pm0.4$, $-3.1\pm0.4$, $-2.4\pm0.4$, and $-1.5\pm0.5$ in each BH mass range.
Our sample shows a wide range of the BH-stellar ratio ($-4.5<\mathrm{log}\ (\MBH/\Mstar)<-0.5$) and the $\MBH$ dependence of the BH-stellar mass ratio implying that no significant correlation exists between $\MBH$ and $\Mstar$ at the observed redshift.
We also obtained slightly positive slopes of $0.21\pm0.11$ and $0.19\pm0.09$ from the linear fitting results in mass ranges $10^7\leq\MBH/\Ms<10^8$ and $10^8\leq\MBH/\Ms<10^9$.
These findings suggest that the BH-stellar mass ratios in our sample with $10^7\leq\MBH/\Ms<10^8$ and $10^8\leq\MBH/\Ms<10^9$ haven't significantly evolved up to their respective redshifts.
Our spectroscopic data sample may suffer from a bias in which less luminous AGNs at high redshift are not involved, as they have not been observed or the S/N ratios of the available spectra are insufficient to measure the broad line width.
Consequently, we expect the slopes to be even flatter when considering the undetected low-mass SMBHs at high redshift.
Due to the small sample of varAGNs with $10^6 \leq \MBH/\Ms<10^7$ and $10^9 \leq \MBH/\Ms<10^{10}$, the redshift evolution in these ranges remains unclear with our varAGN sample alone.


For comparison, we plot the BH-stellar mass ratios of nearby AGNs with $\mathrm{log}\ (\MBH/\Ms)={6.5, 7.5, 8.5}$ taken from the linear regression of \citet{reines2015relations} by the crosses.
Left and right triangles show the BH-stellar mass ratios of spiral galaxies and ETGs with $\mathrm{log}\ (\MBH/\Ms)={6.5, 7.5, 8.5, 9.5}$ from \citet{davis2018blackspiral} and \citet{sahu2019blacketgs}, respectively.
Upward triangles show the BH-bulge mass ratios ($\MBH/\Mbulge$) with $\mathrm{log}\ (\MBH/\Ms)={6.5, 7.5, 8.5, 9.5}$ of elliptical and S0 galaxies in the local universe \citep{kormendy2013coevolution}.
The BH-stellar mass ratio of varAGNs in the BH mass range of $10^6\leq \MBH/\Ms<10^9$ roughly coincides with the ratio of AGNs, Spiral and ETGs observed in the local universe.
On the other hand, varAGNs with $10^9\leq\MBH/\Ms<10^{10}$ show a higher BH-stellar mass ratio than the BH-stellar mass ratio and even the BH-bulge mass ratio in the local universe.
When comparing high-z AGN with our sample, we take the average of all high-z samples within each BH mass range, regardless of their luminosity.
We note that there is a debate on the possible observed bias for the results of high-z luminous quasars.
High-z AGNs with $\MBH=10^{8-9}\ \Ms$ and $\MBH=10^{9-10}\ \Ms$ have been detected by ALMA, showing the high BH-stellar mass ratio on average, $-1.7\pm0.6$ and $-1.9\pm1.0$ (orange and red squares, \citep{izumi2019subaru,pensabene2020alma,izumi2021subaru}).
The BH-stellar mass ratio of our sample in the same BH mass range is similar to these of high-z AGNs.
Recent observations from JWST reveals that AGNs in the low mass range of $\MBH=10^{6-7}\ \Ms$ and $\MBH=10^{7-8}\ \Ms$ at $z=4-6$ also have the high BH-stellar mass ratio on average, $-2.2\pm0.7$ and $-1.3\pm0.5$ (purple and blue diamonds, \citep{harikane2023jwst,maiolino2023jades}).
Compared with our results, we found that the BH-stellar mass ratio for the BH mass range $\MBH=10^{6-8}\ \Ms$ shows significant difference between $z>4$ and $z<3$.
We interpret the behavior as follows; at intermediate redshift or the local universe, the galaxies hosting relatively low-mass SMBH ($\MBH=10^{6-8}\ \Ms$) have already obtain the sufficient total stellar mass, although high-z galaxies are still in an early stage of galaxy formation.

\section{Discussion}

We here discuss the possible interpretation of our results in terms of the co-evolution of SMBHs and their host galaxies.
Our findings reveal the absence of a strong correlation between $\MBH$ and $\Mstar$, although it is unclear whether a correlation exists between $\MBH$ and $\Mbulge$ at intermediate redshift, as observed in the local universe \citep[e.g.,][]{kormendy2013coevolution}.

\subsection{BH-bulge relation for massive SMBH}
Firstly, we consider that the BH-stellar mass ratios with massive SMBHs ($9\leq\MBH/\Ms<10$) at intermediate redshift show large values compared to the local universe. 
In the local universe, most SMBHs in this mass range are typically found in giant elliptical galaxies with small disk contributions.
In Figure \ref{s-ratio-bin}, the red dashed line shows the average value $\mathrm{log}\ (\MBH/\Mstar)=-1.5$, for the 7 objects at $1.5 < z < 3$, which is much larger than the BH-bulge ratio $\mathrm{log}\ (\MBH/\Mbulge)=-2.2$ in the local universe for the same BH mass range \citep[e.g.,][]{kormendy2013coevolution}.
These overmassive SMBHs are rare in the local universe, with only a few reported so far, such as NGC1277, Mrk1216, and PGC032873 \citep{van2012overmassive,yildirim2015mrk,ferre2017twoovermassive}.
These galaxies have extremely high central densities, ended their star formation in the early universe and are primarily composed of red stars.
\citet{ferre2015massive} propose a scenario where the host galaxies of these overmassive SMBHs follow a unique evolutionary path, different from the two-phase growth model typically assumed for massive galaxies. 
After the formation of the SMBH and the galaxy's core, these galaxies bypass the second growth phase, maintaining their original structure without further increases in mass and size.
The overmassive SMBHs at intermediate redshift in our sample could also be formed through this scenario.
However, considering many of these objects will eventually evolve into the established $\MBH-\Mbulge$ relation seen in the local universe, it's necessary to trigger events that increase only the stellar mass.
At high redshift, $z>4$, recent studies using ALMA \citep{izumi2019subaru, pensabene2020alma, izumi2021subaru} have found that the BH-dynamical mass ratios also show the similar large value on average, $\mathrm{log}\ (\MBH/\Mdyn)\ge-2$.
The presence of such large BH-stellar mass ratios not only at high-z but also at intermediate redshift implies that there is a typical growth path of massive SMBHs which is faster than the formation of stars in the associated bulges/spheroids as final products yet to reach the canonical mass ratio in the end.
\subsection{BH-bulge relation for less massive SMBH}
We also discuss the $\MBH-\Mbulge$ relation for less massive SMBHs ($10^7\leq\MBH/\Ms<10^9$) in our sample.
Two possibilities regarding $\MBH-\Mbulge$ relation can be considered to explain the observed result that the BH-stellar mass ratios for the given BH mass do not appear to evolve significantly over the observed redshift range.
The first possibility is that the observed AGNs at intermediate redshift have already developed their present-day structures in terms of the BH, bulge, and disk components, implying that the correlation of the $\MBH-\Mbulge$ relation has already been established in their formation history.
In other words, what we observe as AGN with $10^7\leq\MBH/\Ms<10^9$ at $0.5<z<3$ are nearly the end products of their BH growth and galaxy formation. 
However, this synchronized evolution contrasts with the recent discovery of the overmassive BH in less massive SMBHs range at high redshift, $4<z<7$ \citep[e.g.,][]{harikane2023jwst,maiolino2023jades}. 
Less massive SMBH host galaxies at $z>4$ may be affected by selection bias, potentially leading to an underestimation of their stellar mass.
However, \citet{pacucci2023jwst} reported that the $\MBH-\Mstar$ relation with low-mass SMBH range $\MBH/\Ms<{8.5}$ at $z>4$ deviates significantly from the local relation which is not due to selection effects.
Why, then, do we observe such a difference in the BH-stellar mass ratio between intermediate and high redshift?
\citet{pacucci2024redshift} show the interpretation of the overmassive BHs with the assumption that black hole mass growth is regulated by the quasar's output, and the stellar mass growth is also quenched by it. 
This process works at high redshift when the host galaxies are physically small, and the accretion duty cycle is high, but a runaway process brings the originally overmassive system towards the local BH to stellar mass relation once the condition of suppressing star-formation is broken. 
This may explain the difference of the BH-stellar mass ratio between intermediate and high redshift.

The second possibility is that the BH is overmassive in terms of the BH-bluge mass ratio, as observed at high redshift \citep{harikane2023jwst,maiolino2023jades} or for the more massive SMBH ($>10^9 \Ms$) in our sample, even though this does not appear in the BH-stellar mass ratio. 
BH is being formed in overmassive to their bulges but eventually the surrounding stellar system, probably disks, are formed to show the observed BH-stellar mass ratio. 
Considering that host galaxies have already acquired total stellar mass, these galaxies likely have a larger disk contribution compared to their local universe galaxies in the same BH range.
This redshift evolution of $\MBH-\Mbulge$ relation was also shown by \citet{croton2006evolution} using a semi-analytic simulation.
They argue that if galaxy mergers are the primary drivers for both BH and bulge growth, one should expect the $\MBH-\Mbulge$ relation to evolve with redshift, with a larger black hole mass associated with a given bulge mass at earlier times relative to the present day. In this model, the growth of bulge mass is due to the disruption of the disks in merging events. 
This model with high BH feeding efficiency could explain the high BH-bulge mass ratios observed beyond the local universe.
As an another method to maintain total stellar mass while increasing bulge mass from $z\sim3$ to $z=0$, there is a case that clumps falling into the bulge by dynamical friction.
Since galaxies at intermediate redshift to be much more clumpy than their counterparts in the local universe \citep[e.g.,][]{shibuya2016morphologies}, mass transfer from clumpy disks might be effective for some less massive SMBH hosts, but not for all, inferred from the clumpy galaxy fraction ($<0.7$ at $z=2$ in \citet{shibuya2016morphologies}).

In the context of the redshift evolution of the $\MBH-\Mbulge$ relation for the less massive SMBHs, the host galaxies should have experienced the morphological changes through the merging, disk instability or the clumpy disk transformation.

For future work, using the data from the JWST treasury program, COSMOS-Web, we aim to investigate into the galaxy morphology of our sample, focusing on accurately distinguishing between bulges and disks (including clumps), to explore whether the $\MBH-\Mbulge$ relation evolves with redshift.
\section{Conclusion}
We investigated the relationship between SMBHs and their host galaxies using the variability-selected AGN sample from the HSC-SSP UD survey in the COSMOS field.
Our main results of this work are summarized as follows:
\begin{itemize}
  \item We estimated the $\MBH$ of the deep optical variability-selected AGN sample using the single epoch virial method. Our sample which includes X-ray detected and X-ray undetected AGNs showed wide BH mass range $10^{5.5}<\MBH/\Ms<10^{10}$ and redshift range $0<z<4.2$. We presented that the optical variability-selected AGNs is a powerful tool to investigate the low-mass SMBH at intermediate redshift.
  \item Eddington ratios of X-ray detected and X-ray undetected variability-selected AGNs are converted from hard X-ray luminosity and intrinsic bolometric AGN luminosity of SED results, respectively. The mean Eddington ratio of X-ray detected variability-selected AGNs is $\sim0.1$, showing that of the typical Type1 AGN sample. The mean Eddington ratio of X-ray undetected variability-selected AGNs is $\sim0.06$, showing small acretion rate compared to that of the X-ray detected variability-selected AGNs.
  \item We obtained the $\Mstar$ by separating the AGN and host stellar components with the SED fitting. We found that the $\MBH - \Mstar$ relation in our sample shows little correlation, resulting that the $\MBH/\Mstar$ ratio apparently depends on $\MBH$ up to $z=3.5$. 
  \item The $\MBH/\Mstar$ ratios with $10^6\leq\MBH/\Ms<10^9$ in our variability sample seem to be consistent with the that of AGNs and galaxies observed in the local universe, although the $\MBH/\Mstar$ ratio with $10^9\leq \MBH<10^{10}$ shows significantly high values compared to those in the the local universe.
  \item The $\MBH/\Mstar$ ratio with $10^8\leq\MBH/\Ms<10^{10}$ in our variability sample is roughly consistent with those of AGNs at $z>4$, on the other hand, in the low-mass SMBH ranges of $10^6\leq\MBH/\Ms<10^8$, the $\MBH/\Mstar$ ratio shows small values compared to those at high-z.
  
\end{itemize}
\begin{acknowledgments}
We would like to express our gratitude to Yuki Kimura for his valuable comments and support in developing this work.
This work was supported by JST, the establishment of university fellowships towards the creation of science technology innovation, Grant Number JPMJFS2102.
This publication is based upon work supported by KAKENHI (22K03693) through Japan Society for the Promotion of Science.

\end{acknowledgments}

\bibliography{main}{}
\bibliographystyle{aasjournal}
\appendix

\begin{table*}[ht]
 \centering
 \caption{Filter List used in SED fitting}
  \label{sed-filter}
   \begin{tabular}{ccc}
    \hline \hline
  Telescope   &  Filter Name & Pivot Wavelength [nm]\\
  (1)&(2)&(3)\\
   \hline 
  Chandra(0.5-2 keV) & xray\_boxcar\_0p5to2keV & -\\
  Chandra(0.5-7 keV) & xray\_boxcar\_0p5to7keV & -\\
  Chandra(2-7 keV) & xray\_boxcar\_2to7keV & -\\
  CFHT/Megacam & MCam\_u & 380\\
  CFHT/Megacam & cfht.megacam.u & 382\\
  Subaru/HSC & subaru.hsc.g & 479\\
  Subaru/HSC & subaru.hsc.r & 619\\
  Subaru/HSC & subaru.hsc.i & 767\\
  Subaru/HSC & subaru.hsc.z & 890\\
  Subaru/HSC & subaru.hsc.y & 978\\
  VISTA/VIRCAM & vista.vircam.Y & 1020\\
  VISTA/VIRCAM & vista.vircam.J & 1252\\
  VISTA/VIRCAM & vista.vircam.H & 1643\\
  VISTA/VIRCAM & vista.vircam.Ks & 2152\\
  Subaru/suprime & subaru.suprime.IB427 & 426\\
  Subaru/suprime & subaru.suprime.IB464 & 463\\
  Subaru/suprime & subaru.suprime.IB505 & 506\\
  Subaru/suprime & subaru.suprime.IB574 & 576\\
  Subaru/suprime & subaru.suprime.IB709 & 707\\
  Subaru/suprime & subaru.suprime.IB827 & 824\\
  Subaru/suprime & subaru.suprime.NB711 & 711\\
  Subaru/suprime & subaru.suprime.NB816 & 814\\
  Subaru/suprime & subaru.suprime.B & 444\\
  Subaru/suprime & subaru.suprime.V & 547\\
  Subaru/suprime & subaru.suprime.zpp & 909\\
  Spitzer/irac & spitzer.irac.ch1 & 3556\\
  Spitzer/irac & spitzer.irac.ch2 & 4502\\
  HST/WFC3 & hst.wfc.F814W & 805\\
\hline
\end{tabular}
\begin{tablenotes}
\item[](1)Telescope and Camera (2)Filter name used in CIGALE-v2022.0 (3)Pivot wavelength of the filter
\end{tablenotes}
\end{table*}

\begin{table*}[ht]
 \centering
 \caption{SED CIGALE model parameter}
  \label{CIGALE model}
   \begin{tabular}{ll}
    \hline \hline
   
  Parameter   & Input values \\
   \hline 
   \multicolumn{2}{c}{Star Formation History : sfhdelayed \citep{ciesla2017sfr} }\\
    e-folding time of the main stellar population model in Myr
    & 1000, 3000, 6000, 9000, 12000 \\
   Age of the main stellar population in the galaxy in Myr
    & 1000, 3000, 6000, 9000, 12000  \\
   e-folding time of the late starburst population model in Myr
    & 50  \\
   Age of the late burst in Myr. The precision is 1 Myr
    & 50  \\
   Mass fraction of the late burst population
    & 0.1  \\
    \hline 
    \multicolumn{2}{c}{Stellar Synthesis Population : bc03 \citep{bruzual2003stellar}}\\
    Initial Mass Function
    & Saplpeter \citep{salpeter1955luminosity}   \\
    Metallicity     & 0.02 \\
    \hline 
    \multicolumn{2}{c}{Nebular emission : nebular \citep{inoue2011rest}}\\
    Ionisation parameter     & $-0.2$ \\
    Electron density  & 100 \\
    Fraction of Lyman continuum photons escaping the galaxy & 0 \\
    Fraction of Lyman continuum photons absorbed by dust    & 0 \\
    Line width in km/s &    300.0\\
    \hline
     \multicolumn{2}{c}{Dust attenuation : dustatt$\_$modified$\_$CF00 \citep{charlot2000simple}}\\
    V-band attenuation in the interstellar medium & 0.5, 1, 1.5, 2\\
    Power law slope of the attenuation in the ISM & $-0.7$ \\
    \hline
     \multicolumn{2}{c}{Dust emission : dale2014 \citep{dale2014two}}\\
    Powerlaw slope $dU/dM \propto U^\alpha$ & 1.5, 2.0, 2.5\\
    \hline
     \multicolumn{2}{c}{AGN model : skirtor2016 \citep{stalevski20123d,stalevski2016dust}}\\
    
    Average edge-on optical depth at 9.7 micron & 3, 7, 11 \\
    Inclination  & 0,10, 20, 30, 40, 50\\
    AGN fraction &0.1, 0.2, 0.3, 0.4, 0.5, 0.6, 0.7, 0.8, 0.9\\
    Extinction law of the polar dust & SMC \\
    E(B-V) for the extinction in the polar direction in magnitudes
    & 0, 0.03, 0.1\\
    \hline
     \multicolumn{2}{c}{X-ray : from AGN, galaxy \citep{stalevski2016dust}}\\
    Photon index $\Gamma$ of the AGN intrinsic X-ray spectrum & 1.4, 1.6, 1.8\\
   Power law slope connecting $L_\nu$ at rest frame $2500\AAA$ and 2 keV : $\alpha_\mathrm{{ox}}$ & $-1.8$, $-1.7$, $-1.6$, $-1.5$, $-1.4$, $-1.3$\\
   Maximum allowed deviation of $\alpha_\mathrm{{ox}}$ from the empirical $\alpha_\mathrm{{ox}}-L_\nu$ 2500 \AA  & 0.2 \\
   \hline
    
  \end{tabular}
 \end{table*}


\clearpage
\begin{table*}[t]
 \centering
 \caption{The physical parameters of variability-selected AGNs with estimated $\MBH$}
  \label{bhmass-table1}
   \begin{tabular}{cccccccccc}
    \hline \hline
    varAGN ID & R.A.(J2000) & Dec.(J2000) & $i$-band & Redshift & $FWHM$ & $\MBH$ & $\Lbol$ & S/N & line \\
    \hline
      & [deg] & [deg] & [mag] &  & [$10^3\ \mathrm{km\ s^{-1}}$] & [$\Ms$] & [$\mathrm{erg\ s^{-1}}$] &  &  \\
    \hline
    (1) & (2) & (3) & (4) & (5) & (6) & (7) & (8) & (9) & (10) \\
\hline
&&&&X-det varAGN&&&&\\

1 & 150.74386 & 2.20245 & 22.71 & 1.618 & 6.4 $\pm$ 1.17 & 8.6 $\pm$ 1.08 & 45.3 & 6.63 & 2 \\
2 & 150.73557 & 2.19957 & 20.36 & 3.497 & 5.6 $\pm$ 0.03 & 8.8 $\pm$ 0.16 & 45.9 & 4.76 & 2 \\
7 & 150.78838 & 2.34399 & 19.82 & 1.964 & 4.6 $\pm$ 0.00 & 8.5 $\pm$ 0.07 & 45.8 & 7.46 & 2 \\
8 & 150.71511 & 2.48483 & 19.20 & 2.003 & 4.4 $\pm$ 0.00 & 8.5 $\pm$ 0.04 & 45.8 & 8.95 & 2 \\
12 & 150.65294 & 1.99685 & 19.52 & 1.518 & 3.0 $\pm$ 0.03 & 7.9 $\pm$ 0.19 & 45.1 & 8.03 & 1 \\
15 & 150.61224 & 1.99442 & 21.40 & 1.609 & 3.1 $\pm$ 0.16 & 7.8 $\pm$ 0.40 & 45.0 & 6.42 & 1 \\
16 & 150.57489 & 1.97677 & 20.98 & 1.541 & 4.9 $\pm$ 0.01 & 8.5 $\pm$ 0.08 & 45.5 & 11.98 & 1 \\

\hline
\hline
&&&&X-undet varAGN&&&&\\
56 & 150.42931 & 1.82558 & 21.74 & 2.084 & 5.9 $\pm$ 0.17 & 8.4 $\pm$ 0.42 & 45.0 & 0 & 2 \\
108 & 150.36187 & 2.77911 & 22.54 & 2.537 & 6.5 $\pm$ 0.51 & 8.5 $\pm$ 0.71 & 45.2 & 0 & 2 \\
113 & 150.30622 & 1.87518 & 23.31 & 1.755 & 2.2 $\pm$ 0.08 & 7.6 $\pm$ 0.28 & 45.2 & 0 & 1 \\
\hline
\end{tabular}
\begin{tablenotes}
\item[](1)(2)(3)(4) ID, Right Ascension, Declination and $i$-band AB magnitude are taken from \citet{kimura_properties_2020}.  (5) spectroscopic redshift (6)(7) the FWHM of the broad line, mass of the SMBH (8) bolometric luminosity converted from hard X-ray luminosity and SED fitting result for X-ray detected AGNs and  X-ray undetected AGNs respectively. (9) Signal to noise ratio of hard X-ray in the Chandra COSMOS Legacy Survey \citep{laigle2016cosmos2015,marchesi2016chandra} (10) the label of the broad lines we use to estimate $\MBH$, 1 :$\Mgtwo$, 2: $\Cfour$, 3: $\Hb$, 4: $\Ha$. The full data can be available in the machine-readable version.
\end{tablenotes}
\end{table*}

\begin{table*}[t]
 \centering
 \caption{Summary of our results}
  \label{bh-ste-table}
   \begin{tabular}{ccccc}
\hline
\hline
varAGN ID & Redshift & $\MBH$ & $\Mstar$ & $\chi_\mathrm{red}$ \\
\hline
 &  & $[\Ms]$ & $[\Ms]$ & \\
\hline
(1) & (2) & (3) & (4) & (5) \\
\hline

1 & 1.618 & 8.6 $\pm$ 1.08 & 11.0 $\pm$ 0.71 & 2.3 \\
2 & 3.497 & 8.8 $\pm$ 0.16 & 11.5 $\pm$ 0.49 & 1.5 \\
7 & 1.964 & 8.5 $\pm$ 0.07 & 10.3 $\pm$ 0.15 & 3.4 \\
12 & 1.518 & 7.9 $\pm$ 0.19 & 10.8 $\pm$ 0.21 & 1.9 \\

  \hline
\end{tabular}
\begin{tablenotes}
\item[](1) varAGN IDs are taken from \citet{kimura_properties_2020}.  (2) Spectroscopic redshift (3) the mass of the SMBHs (4) the total stellar mass of the host galaxies (5) reduced chi-square in SED fitting results. The full data can be available in the machine-readable version.
\end{tablenotes}
\end{table*}

\end{document}